\documentclass[screen]{acmart}
\AtBeginDocument{%
  \providecommand\BibTeX{{%
    \normalfont B\kern-0.5em{\scshape i\kern-0.25em b}\kern-0.8em\TeX}}}

\setcopyright{acmlicensed}
\copyrightyear{2024}
\acmYear{2024}
\acmDOI{XXXXXXX.XXXXXXX}

\acmConference{CHI 2024: ACM CHI conference on Human Factors in Computing Systems}{May 11--16, 2024}{Honolulu, HI}

\usepackage{graphicx}
\usepackage{caption}
\usepackage{subcaption}

\begin{document}

\title[Tinker or Transfer?]{Tinker or Transfer? A Tale of Two Techniques in Teaching Visualization}

\author{Murtaza Ali}
\authornote{Both authors contributed equally to this research.}
\email{mali53@uw.edu}
\orcid{0000-0002-2998-3165}
\author{Adam Hyland}
\authornotemark[1]
\email{achyland@uw.edu}
\orcid{0000-0002-2570-6689}
\affiliation{%
  \institution{University of Washington}
  \city{Seattle}
  \state{Washington}
  \country{USA}
}

\renewcommand{\shortauthors}{Ali and Hyland}

\begin{abstract}
In education there exists a tension between two modes of learning: traditional lecture-based instruction and more tinkering-based creative learning. In this paper, we outline our efforts as two Ph.D. students (who are skilled in visualization but are not, importantly, professionally trained visualization experts) to implement creative learning activities in an information visualization course in our home department. We describe our motivation for doing so, and how what began out of necessity turned into an endeavor whose utility we strongly believe in. In implementing these activities, we received largely positive reviews from students, along with constructive feedback which helped us iteratively improve the activities. Finally, we also detail our future plans for turning this work into a formal design inquiry with students to build a new class centered entirely around creative learning.
\end{abstract}



\received{16 April 2024}

\maketitle

\section{Introduction}
\begin{quote}
\textbf{\textit{``We are not formally trained visualization experts. How do we teach this class?''}}
\end{quote}

As two junior PhD students given the opportunity to teach an information visualization course in the University of Washington’s Human-Centered Design and Engineering Department, this is the central question we grappled with. It is not that we were unqualified, but we were starkly aware of the fact that students could have just as easily enrolled in a similar course with pioneers like Jeffrey Heer and Jessica Hullman. We felt anxious, knowing we could not teach visualization with their same authority, especially in a course structured around the expectation that an expert instructor would transfer (as Lave uses the term ``transfer theory of learning'' in \textit{Cognition in Practice} \cite{lave_cognition_2009}) general visualization principles to students to be applied to a variety of problems and contexts.

We struggled with this for multiple reasons. First, visualization is an astonishingly broad field which incorporates human factors, design, physics, anatomy (the retina!), and data science–just to name a few foundational disciplines. We are not experts in these fields–neither of the authors can, for instance, hold court on the retina for more than a few minutes. Second, such as we can claim qualification in visualization, the transfer theory does not represent how we learned what we have. We learned through muddling, tinkering, failing, and iterating–through play, and the creative application of practical lessons. Finally, our students come from a wide range of backgrounds, from mechanical engineering to biology to user research; incorporating diverse sources of expertise through project-based learning is a core value of our department. Transmission of principles from us as instructors who learned non-traditionally to students who might bring a range of novel insights given the chance did not quite fit with this value. 

Knowing the above, we wanted to expose students to the topic as we were. This means we have had to expose ourselves to the possibility that our efforts might not ‘come off’ (to borrow J.L. Austin’s phrasing \cite{austin_how_2009}) because both the possibility of failure and the outcome of play are unknown at the outset. It was not our aim to disavow the instruction of general principles of visualization. Furthermore, the instructors we learned from and modeled the class after are deeply invested in student-led, practical learning. We aim to make an evolutionary change, not a large disruption. Our interventions are meant to be partial–it is not our assertion that the transfer theory of learning does not work, rather that it exists in concert (more or, in a classroom, perhaps less) with other modes of learning. 

We started this journey out of necessity, realizing this was the most effective way we could teach the course. But each time we taught the course, we became more convinced that the strategies we were using would help anyone teach visualization better–that teaching visualization in this way helps students learn the requisite concepts and leave with the practical ability to build visualizations. This paper details our work to date in the classroom with creative learning activities, the student response, and our plans to map these activities to learning assessments in the future. Our aim is to provide concrete examples of creative learning activities in a visualization classroom so instructors who wish to adopt them will have models on which to base their efforts and instructors who already include activities like these in their lesson plans will be prompted to share them in the literature. 

\section{Background and Related Work}
\subsection{Creating Learning in Visualization}
While there is a wide literature regarding how technological advances in visualization tools can enable creative learning practices (Kamath and Kamat offer, for instance, ``Cost-Effective 3D Stereo Visualization for Creative Learning'' \cite{khosrow-pour_dba_advanced_2019}) or how the cognitive benefits of visualizing data or processes can allow for improvements in understanding of scientific concepts \cite{gilbert_visualization_2005, vavra_visualization_2011}, there is limited work we have found so far addressing or connecting creative learning to teaching visualization. The most salient example of such work is a 2022 paper by Roberts et al. reflecting on integrating creative activities into learning visualization, ``Reflections and considerations on running creative visualization learning activities'' \cite{roberts_reflections_2022}. This paper, while valuable, is itself evidence of the paucity of literature on the subject–the authors met at a workshop and by happenstance discovered their individual creative learning practices contained deep commonalities. They offer a collection of best practices for instructors to follow when designing activities like ours; however we submit the field needs not just best practices but examples tied to assessment. We ought to talk about and circulate activities as well as advice.

While research into visualization tools circulates well in the academy, much of the work of teaching is invisible. Teaching practice is recorded and instantiated in syllabi and lesson plans \cite{aydogdu_analysis_2023}, which often circulate only informally, making it difficult for new and even seasoned instructors to learn from teaching practices of others. This is beginning to change, with many of the same participants in the workshop which spawned Roberts et al. 2022 writing a call to action that identifies a prime opportunity to build a community of visualization educators \cite{bach_challenges_2023}.

Creative learning is a broad field with many contributing theories; much of our own perspective is influenced by Seymour Papert’s constructionism (c.f. especially \textit{Mindstorms} \cite{papert_mindstorms_1982}) and subsequent ideas. We use the term "creative learning" to refer to a body of practices which are centered on student exploration of an area through experimentation and play. Papert's constructionism was a pedagological approach connected to bringing students into the complex world of geometry and logic through play with machines. Here we adapt this as Roberts et al. do to play with visualization as a view into the world of visualization literacy. As with computing, creative learning is a means to build a working understanding of visualization, which students may deploy and adapt to be informed creators and consumers of increasingly complex visual representations.

There are challenges in porting over creative learning practices to the teaching of visualization. If we look at Resnick et al.'s 1996 “Pianos, not Stereos,” a core goal of creating a computational construction kit is exposing the inner workings in a way that tinkering with them and observing them will connect learners to insights about their function and capability \cite{resnick_pianos_1996}. Many pianos (at least those which cost thousands of dollars) can be inspected during use and while you may not always want to stick your finger in one, it is likely to be a less shocking experiment than doing so with a stereo. However, it is hard to open the hood on visualization software and show anything similar to a hammer falling on a wire. Even open source tools which allow for total inspection (no black boxes) are more like stereos than pianos--they are difficult to manipulate without breaking. As we teach visualization in 2024, we must bear in mind that many of the practices of visualization have been cemented and embedded in the toolspace (see here Card and MacKinlay 1997 \cite{card_structure_1997}) already.

We teach two visualization courses, an upper level undergraduate course which requires a prerequisite introductory Python course and a master’s course which has no requirements aside from graduate standing. As such, the student pool has a wildly varying set of skills to be applied to the problem space. We teach designers, ethnographers, engineers, and programmers to visualize, not only by applying visual principles, but ensuring that students are able to bring their expertise to bear on the cooperative problem of visual design. Turkle and Papert’s 1990 work on epistemological pluralism (``Epistemological Pluralism: Styles and Voices within the Computer Culture,'' \cite{turkle_epistemological_1990}), observes that while many students did use traditional, abstract approaches to programming, others used equally valid approaches ``more reminiscent of a painter than a logician.'' Porting this over to a visualization context, we similarly want to encourage students to bring their diverse points of view to contribute a cohesive whole. We find that they are able to do so very effectively through these creative learning activities.

We have also found more uncomplicated parallels. In ``It looks like fun, but are they learning?'' Petrich et al. discuss a common response to the introduction of creative learning practices into what is often an evaluative landscape: the presence of fun and chaos can often be a sign that a certain mode of instruction has gone astray \cite{petrich_it_2013}. Part of our aim is to exhort (sometimes by example) students to ``play around with'' or ``tinker'' with the visualization tools available. This represents a reasonably significant break with practice in other visualization courses in our university, which take the time we devote to activities and use it to convey knowledge about visualization principles. We might, a bit cheekily, ask ourselves, ``It looks like viz, but are they learning?'' considering that only a fraction of our exercises are explicit problem/solution pairs. Coupling these activities to evaluative measures directly is an ongoing process and one which we approach in the discussion and future work sections. 

\section{Methodology}
\subsection{Class Structure and Evaluation}
We introduced 15-minute activities at the start of each class designed to give the students an opportunity to tinker with visualization concepts.

These 15-minute activities had minimal structure the first time we incorporated them into our course in Spring 2023. They related sometimes to the topic of the day, other times to visualization in general. We often came up with them on the fly as warm-up activities to get students ready for the class session. However, seeing their effectiveness, we made a concerted effort to structure and formalize them for the next run of the class in Autumn 2023. Each activity is delineated by three features:

\begin{itemize}
  \item A \textbf{description} of the activity itself.
  \item A list of \textbf{materials} needed to successfully run the activity.
  \item A detailed explanation of the \textbf{rationale} behind the activity.
\end{itemize}

In formalizing these activities, we had one overarching goal from the design perspective. Each activity is designed to correspond with the day’s content, to frame students’ thinking so that they are already implicitly using conceptual knowledge before being introduced to it formally \cite{katajavuori_significance_2006}.

Below, we provide three illustrative examples of the 15-minute activities before delving into the outcomes. We reproduce them here as they are described in our class materials, with only minor changes for clarity. The descriptions below are used as a guide for instructors and not shared directly with students.

\subsection{Activity Examples}
\subsubsection{Activity 1: Introduction}
\begin{itemize}
  \item \textbf{Activity}: Students are presented with a set of 10 cups arranged in a triangle. Their task is to rearrange a maximum of three cups in order to flip the triangle upside down.
  \item \textbf{Materials}: Colorful shot cups, pen and paper for them to work.
  \item \textbf{Rationale}: At its core, data visualization is just a way of arranging data. Its effectiveness is dependent on how well the data visualizer plans and arranges said data. How can different arrangements convey the same information? How are some arrangements better than others? Is there a single best arrangement?
\end{itemize}

\subsubsection{Activity 2: How Do We Encode Data Well?}
\begin{itemize}
  \item \textbf{Activity}: In groups, design and construct (on paper) a visualization which explains multiplication tables to a class of third graders. Take into consideration the needs and abilities of the audience. Be prepared to justify the visual choices made as well as answer the following question: How did traditional design and implementation approaches need to be changed to accommodate this specific audience? Any visualization is acceptable, with the exception of the traditional tabular format in which multiplication tables are represented.
  \item \textbf{Materials}: Pen and paper.
  \item \textbf{Rationale}: Data visualization is \textit{hard}. Figuring out how to best represent data in visual form is no easy task–this is the foundational idea of choosing visual encoding channels. From time to time, these activities will simply aim to give practice with quickly prototyping a visualization for a specific use case and justifying the choices made. There will often be a thematic question for the activity.
\end{itemize}

\subsubsection{Activity 3: Fibonacci, But No Trees!}
\begin{itemize}
  \item \textbf{Activity}: In groups, develop a visualization (this can be on paper or in a visualization software) to explain the Fibonacci sequence to someone who is having trouble understanding it. Any visualization is acceptable, with the exception of a tree-like visualization which traces the sequence’s development.
  \item \textbf{Materials}: Pen and paper.
  \item \textbf{Rationale}: This is another instance of simply giving practice thinking about how one might visualize different types of data, similar to the multiplication tables one from some days prior.
\end{itemize}

\begin{center}
\begin{figure}[ht!]
  \begin{subfigure}[b]{0.4\textwidth}
    \includegraphics[width=\textwidth]{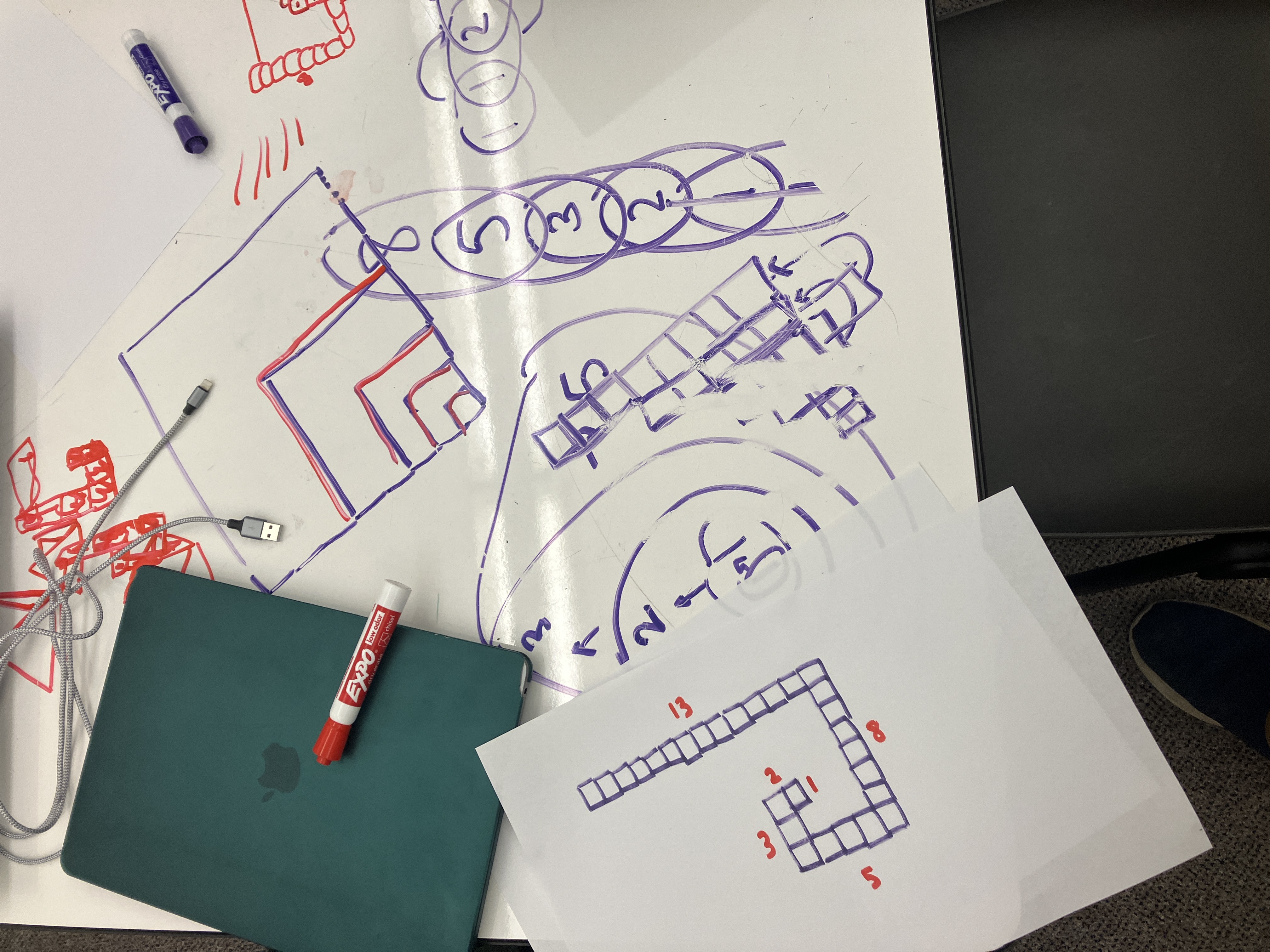}
    \caption{In-progress sketches for the Fibonacci exercise}
    \label{fig:f1}
  \end{subfigure}
  \hspace{3em}
  \begin{subfigure}[b]{0.4\textwidth}
    \includegraphics[width=\textwidth]{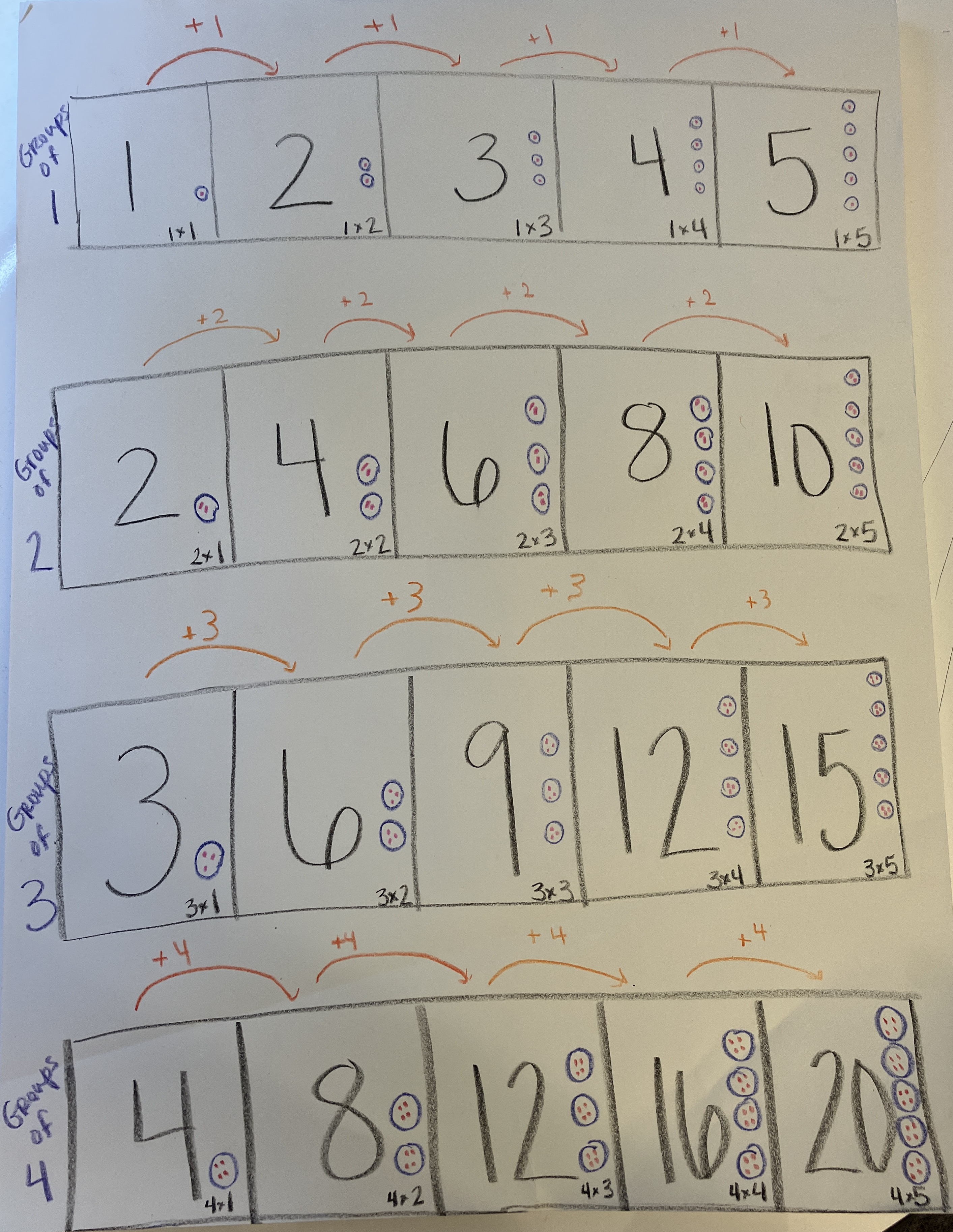}
    \caption{A didactic multiplication visualization}
    \label{fig:f2}
  \end{subfigure}
  \caption{Example Student sketches and outputs for two of the exercises}
\end{figure}
\end{center}

\subsection{Three Thoughts Regarding the Activities}
All of our activities are collaborative and involve student discussion and ideation, and are deliberately such. Mary Eleanor Spear, an early data visualization pioneer, argued \cite{spear_practical_1969} that data visualization required three roles in concert: a communicator, a data analyst and a drafter. The roles themselves have changed and are much fuzzier than before but appreciation for the importance of interaction and collaboration in visualization has grown considerably. 

They are also designed to bring in some skills outside visualization (project management, mathematics, puzzle solving). This too was intentional, aimed at fulfilling our earlier mentioned aim of providing students with an outlet to learn visualization while simultaneously applying their own unique perspectives and backgrounds. 

Our activities set strong constraints on the activity. The task always defines a precise goal and often gives students guidance on how to correctly approach it. Visualizing and conceptualizing multiplication through tables is deeply ingrained in many students,  and a tabular arrangement can be easily designed with little thought to scope and context–constraining students by excluding this as a possibility invites them to think about the problem in unexpected ways. Similarly, many visualizations of the Fibonacci sequence demonstrate its recursive qualities with a tree structure–students with a computer science background might want to resort to these kinds of representations and so miss out on how to explicitly show recursion and scale visually. This is in line with the arguments set out by Roberts et al. in their reflections paper: Constraints are good and can foster creativity!

\section{Outcomes}
After running these activities for the first time in Spring 2023, we evaluated them with a class survey. We received a total of 22 respondents for the survey, with respondents reporting the following high-level statistics:
\begin{itemize}
    \item On a Likert scale of 1-5. 78\% of respondents reported a 4 or higher when asked, ``Do you feel you had the opportunity to learn creatively in this class?''
    \item 69\% reported a 4 or higher when asked, ``How helpful for your learning were the hopefully fun, slightly odd 15-minute activities at the start of each class (making a visualization story out of stickers, puzzles with the colorful shot cups, identifying a rock and developing a related visualization, etc.)?''
\end{itemize}

We also included some free-response questions to gauge specific feedback on the respective activities from the class. By and large, the responses were promising and seemed to indicate we made progress in meeting our initial goals. Referencing the triangle activity, one student stated, ``One of the first class sessions: the activity with the 6 cups arranged in a triangle and having to flip the entire shape around by moving only three of the cups. It was a really good way to wake up my brain and it also got me thinking about different ways data can be arranged.'' For the multiplication activity, another noted, ``I think the exercise where we tried to visualize a math concept to an 8-year-old was really challenging but also meaningful in warming up to understanding how visualizations can work to teach.'' And paying an ode to our friend Fibonacci, one also stated, ``Visualizing the Fibonacci sequence was cool and pushed my thinking when it came to making something that could be easily digested by people.''

Discussions in class surrounding the activities also allowed students to reflect on how different groups of students approached each problem. An advantage of asking students from different backgrounds to build a visualization of an abstract mathematical concept like the Fibonacci sequence is that some undertook methods that relied on iteration and recursion which are often more easily surfaced by those with some experience, but others approached them from divergent viewpoints. As students came together to see each of the visualizations, they could reflect on the various solutions. Sometimes this caused students who had not been thinking about recursion to wonder how they might have visualized that concept but just as often it allowed for students who approached the task from a standard viewpoint to see how an artist or a biologist might do so. 

Finally, there were also a subset of responses that mentioned the activities themselves didn’t necessarily help them learn visualization concepts, but were really helpful as mental warm ups. From a holistic learning and tinkering standpoint, we consider this to be more success than failure. One particularly useful comment from these responses was a student who expressed that they had difficulty connecting the activities to the course content. This feedback spurred us to revise and formalize the activities for the next course run, paying particular attention to drawing connections with course content, as we mentioned above. As an additional minor note, it is also encouraging that the activities stood out enough for the students to actually remember them in detail weeks after the fact.

\section{Discussion}
In their original paper on running creative visualization learning activities, Roberts et al. provide several best practices for implementing such activities in a classroom:
\begin{itemize}
    \item Introduce composition through conceptual and concrete building blocks
    \item Make it fun
    \item Creativity in context and providing freedom to encourage creativity
    \item Different learning materials facilitate learning
    \item Debate and discussions
    \item The need for a sketch
    \item Step outside your comfort zone as a catalyst for creativity
    \item Constraints as scaffolding
\end{itemize}

These guidelines provide an excellent framework for designing such activities, and though we did not explicitly review this paper when designing our activities, we find that our activities overall follow similar patterns to the author recommendations. We could boldly congratulate ourselves on our good judgment and suggest that we have an understanding of such activities, but a better lesson to draw from Roberts et al. is the likelihood that many visualization teachers have a partial grasp of these principles. It is in sharing them and work which exemplifies them to peers that we hope to contribute. 

However, we acknowledge the potential issue here with replication \cite{cacioppo_social_2015}. Will everyone want to replicate the Fibonacci or multiplication tables activity? One of the authors has a background in mathematics which directly influenced his interest in designing such activities. Would instructors with different subject matter expertise prefer different activities? Taking an optimistic approach, we maintain that even if an instructor chose to change the specifics of the activities to better suit their background, the nature and outcomes of the activities we present are replicable. Namely, it is possible to prime students to learn visualization concepts in a wide variety of ways \cite{nolan_teaching_2016}.

There is also the issue of time. These are 15-minute activities in a 110-minute class session. Accounting for breaks and idle time, this takes up a significant chunk of class time–nearly 15\%! Any instructor is familiar with the frustration of trying to fit an immense amount of material into limited time, and including these activities would further necessitate the cutting down of potentially useful content. We recognize this, but we also maintain that because hands-on learning is so valuable in visualization courses \cite{hudiburgh_data_2020}, we feel the benefits outweigh the drawbacks. Furthermore, if designed correctly, these activities can be equally effective in transmitting this content.

Beyond prompting students to approach the problem of creating visualizations differently, these activities can help students gain some new perspectives on visualization literacy. For the multiplication example, asking students to undertake a constrained audience analysis caused some to think about where and when commonly used encodings like color help them understand and internalize visualizations of seemingly simple topics. For our cups example, offered very early in the course, students are invited to talk about the role of organization and presentation in understanding of genre conventions around visualization. We find that these discussions help students navigate more complex and articulated visualizations.

\section{Conclusion and Future Work}
Students like these activities and report that they are fun to participate in. We like these activities and find ourselves looking forward to each new class session with excitement because of them. The activities themselves--which ask students to design under constraint, undertake audience analysis, and transform a problem in a short number of steps--feel like they ought to be valuable. But just because students and teachers like something that ought to be valuable to learning doesn’t make it so.

While we have issued surveys and questioned students about their assessment of the value of the exercises, questions from an instructor such as ``Was this activity helpful to learning?'' can elicit positive responses equally well from students with differing opinions on the material. Our survey responses and informal discussions with students gave us enough confidence to proceed to the in-progress work of assessing the impact of these activities on design and visualization literacy. But from them alone, there is only so much we can learn.

One approach to assuage these concerns might be evaluation via some form of assessment, a traditionally engineering-based approach. Rather than move immediately to evaluation, we first want to adopt a design-oriented approach, as we think it important to probe the student learning space as much as possible before even attempting something so bold as objective evaluation. Our future plans include taking these activities, and creative learning ideas more broadly, and developing them into a new course entirely. We aim to conduct a design inquiry in the early stages of developing this course \cite{sandoval_conjecture_2014}, incorporating students into the design process itself. Some of the problems of replicating creative learning stem from idiosyncratic instructors–the two of us do not scale–and while assessment can identify a failure of replication, co-design with students can work to mitigate them in the first place.

We hope to share out and subject to scrutiny classroom activities which are still evolving, partially to gain feedback which will allow us to better teach visualization but also to prompt others in the community to share lesson plans, activities, and artifacts from their creative visualization teaching practice. The more of us that do so, the less anxiety future junior scholars will have in teaching visualization as they learned it, by tinkering and practical exploration.

\section*{Acknowledgements}

The authors wish to thank Carly Minow, Georgia Kenderova, and Matthew Pedraja for comments and advice as well as Prerna Rao for assistance with formatting and copyediting. We would also like to acknowledge Dr. Sayamindu Dasgupta for introducing us to the world of creative learning and supporting us in our attempts at implementation. All errors are our own.

\bibliographystyle{ACM-Reference-Format}
\bibliography{chi-cl-bib}

\end{document}